\documentclass[aps,prab,twocolumn,superscriptaddress]{revtex4-2} 

\usepackage{lineno,hyperref}%
\usepackage{booktabs}

\usepackage{graphicx}
\usepackage{soul}
\usepackage{color}
\usepackage{textgreek}
\usepackage{placeins}
\usepackage{amssymb}
\usepackage[dvipsnames]{xcolor}

\newcommand{\mr}[1]{\mathrm{#1}}

\newcommand{\fpe}{f_{\mr{pe}}}
\newcommand{\fzero}{f_{\mr{pe0}}}
\newcommand{\fmod}{f_{\mr{mod}}}
\newcommand{\lzero}{\lambda_{\mr{pe0}}}
\newcommand{\lpe}{\lambda_{\mr{pe}}}
\newcommand{\um}{\textmu m}

\begin{document}



\title{Simulation and Experimental Study of Proton Bunch Self-Modulation in Plasma with Linear Density Gradients}



\author{P.I.~Morales~Guzm\'{a}n}
\email[]{pmorales@mpp.mpg.de}
\affiliation{Max Planck Institute for Physics, Munich, Germany}
\author{P.~Muggli}
\affiliation{Max Planck Institute for Physics, Munich, Germany}

\author{R.~Agnello}
\affiliation{Ecole Polytechnique Federale de Lausanne (EPFL), Swiss Plasma Center (SPC), Lausanne, Switzerland}
\author{C.C.~Ahdida}
\affiliation{CERN, Geneva, Switzerland}
\author{M.~Aladi}
\affiliation{Wigner Research Center for Physics, Budapest, Hungary}
\author{M.C.~Amoedo Goncalves}
\affiliation{CERN, Geneva, Switzerland}
\author{Y.~Andrebe}
\affiliation{Ecole Polytechnique Federale de Lausanne (EPFL), Swiss Plasma Center (SPC), Lausanne, Switzerland}
\author{O.~Apsimon}
\affiliation{University of Liverpool, Liverpool, UK}
\affiliation{Cockcroft Institute, Daresbury, UK}
\author{R.~Apsimon}
\affiliation{Cockcroft Institute, Daresbury, UK} 
\affiliation{Lancaster University, Lancaster, UK}
\author{A.-M.~Bachmann}
\affiliation{Max Planck Institute for Physics, Munich, Germany}
\author{M.A.~Baistrukov}
\affiliation{Budker Institute of Nuclear Physics SB RAS, Novosibirsk, Russia}
\affiliation{Novosibirsk State University, Novosibirsk, Russia}
\author{F.~Batsch}
\affiliation{Max Planck Institute for Physics, Munich, Germany}
\author{M.~Bergamaschi}
\affiliation{Max Planck Institute for Physics, Munich, Germany}
\author{P.~Blanchard}
\affiliation{Ecole Polytechnique Federale de Lausanne (EPFL), Swiss Plasma Center (SPC), Lausanne, Switzerland}
\author{F.~Braunm\"{u}ller}
\affiliation{Max Planck Institute for Physics, Munich, Germany}
\author{P.N.~Burrows}
\affiliation{John Adams Institute, Oxford University, Oxford, UK}
\author{B.~Buttensch{\"o}n}
\affiliation{Max Planck Institute for Plasma Physics, Greifswald, Germany}
\author{A.~Caldwell}
\affiliation{Max Planck Institute for Physics, Munich, Germany}
\author{J.~Chappell}
\affiliation{UCL, London, UK}
\author{E.~Chevallay}
\affiliation{CERN, Geneva, Switzerland}
\author{M.~Chung}
\affiliation{UNIST, Ulsan, Republic of Korea}
\author{D.A.~Cooke}
\affiliation{UCL, London, UK}
\author{H.~Damerau}
\affiliation{CERN, Geneva, Switzerland}
\author{C.~Davut}
\affiliation{Cockcroft Institute, Daresbury, UK} 
\affiliation{University of Manchester, Manchester, UK}
\author{G.~Demeter}
\affiliation{Wigner Research Center for Physics, Budapest, Hungary}
\author{A.~Dexter}
\affiliation{Cockcroft Institute, Daresbury, UK} 
\affiliation{Lancaster University, Lancaster, UK}
\author{S.~Doebert}
\affiliation{CERN, Geneva, Switzerland}
\author{J.~Farmer}
\affiliation{CERN, Geneva, Switzerland}
\affiliation{Max Planck Institute for Physics, Munich, Germany}
\author{A.~Fasoli}
\affiliation{Ecole Polytechnique Federale de Lausanne (EPFL), Swiss Plasma Center (SPC), Lausanne, Switzerland}
\author{V.N.~Fedosseev}
\affiliation{CERN, Geneva, Switzerland}
\author{R.~Fiorito}
\affiliation{Cockcroft Institute, Daresbury, UK} 
\affiliation{University of Liverpool, Liverpool, UK}
\author{R.A.~Fonseca}
\affiliation{ISCTE - Instituto Universit\'{e}ario de Lisboa, Portugal} 
\affiliation{GoLP/Instituto de Plasmas e Fus\~{a}o Nuclear, Instituto Superior T\'{e}cnico, Universidade de Lisboa, Lisbon, Portugal}
\author{I.~Furno}
\affiliation{Ecole Polytechnique Federale de Lausanne (EPFL), Swiss Plasma Center (SPC), Lausanne, Switzerland}
\author{S.~Gessner}
\affiliation{CERN, Geneva, Switzerland}
\affiliation{SLAC, Menlo Park, CA, USA}
\author{A.A.~Gorn}
\affiliation{Budker Institute of Nuclear Physics SB RAS, Novosibirsk, Russia}
\affiliation{Novosibirsk State University, Novosibirsk, Russia}
\author{E.~Granados}
\affiliation{CERN, Geneva, Switzerland}
\author{M.~Granetzny}
\affiliation{University of Wisconsin, Madison, Wisconsin, USA}
\author{T.~Graubner}
\affiliation{Philipps-Universit{\"a}t Marburg, Marburg, Germany}
\author{O.~Grulke}
\affiliation{Max Planck Institute for Plasma Physics, Greifswald, Germany}
\affiliation{Technical University of Denmark, Lyngby, Denmark}
\author{E.~Gschwendtner}
\affiliation{CERN, Geneva, Switzerland}
\author{E.D.~Guran}
\affiliation{CERN, Geneva, Switzerland}
\author{V.~Hafych}
\affiliation{Max Planck Institute for Physics, Munich, Germany}
\author{J.R.~Henderson}
\affiliation{Cockcroft Institute, Daresbury, UK}
\affiliation{Accelerator Science and Technology Centre, ASTeC, STFC Daresbury Laboratory, Warrington, UK}
\author{M.~H{\"u}ther}
\affiliation{Max Planck Institute for Physics, Munich, Germany}
\author{M.{\'A}.~Kedves}
\affiliation{Wigner Research Center for Physics, Budapest, Hungary}
\author{V.~Khudyakov}
\affiliation{Heinrich-Heine-Universit{\"a}t D{\"u}sseldorf, D{\"u}sseldorf, Germany}
\affiliation{Budker Institute of Nuclear Physics SB RAS, Novosibirsk, Russia}
\author{S.-Y.~Kim}
\affiliation{UNIST, Ulsan, Republic of Korea}
\affiliation{CERN, Geneva, Switzerland}
\author{F.~Kraus}
\affiliation{Philipps-Universit{\"a}t Marburg, Marburg, Germany}
\author{M.~Krupa}
\affiliation{CERN, Geneva, Switzerland}
\author{T.~Lefevre}
\affiliation{CERN, Geneva, Switzerland}
\author{L.~Liang}
\affiliation{Cockcroft Institute, Daresbury, UK}
\affiliation{University of Manchester, Manchester, UK}
\author{N.~Lopes}
\affiliation{GoLP/Instituto de Plasmas e Fus\~{a}o Nuclear, Instituto Superior T\'{e}cnico, Universidade de Lisboa, Lisbon, Portugal}
\author{K.V.~Lotov}
\affiliation{Budker Institute of Nuclear Physics SB RAS, Novosibirsk, Russia}
\affiliation{Novosibirsk State University, Novosibirsk, Russia}
\author{M.~Martyanov}
\affiliation{Institute of Applied Physics RAS, Nizhny Novgorod, Russia} 
\author{S.~Mazzoni}
\affiliation{CERN, Geneva, Switzerland}
\author{D.~Medina~Godoy} 
\affiliation{CERN, Geneva, Switzerland}
\author{J.T.~Moody}
\affiliation{Max Planck Institute for Physics, Munich, Germany}
\author{K.~Moon}
\affiliation{UNIST, Ulsan, Republic of Korea}
\author{M.~Moreira}
\affiliation{GoLP/Instituto de Plasmas e Fus\~{a}o Nuclear, Instituto Superior T\'{e}cnico, Universidade de Lisboa, Lisbon, Portugal}
\author{T.~Nechaeva}
\affiliation{Max Planck Institute for Physics, Munich, Germany}
\author{E.~Nowak}
\affiliation{CERN, Geneva, Switzerland}
\author{C.~Pakuza}
\affiliation{John Adams Institute, Oxford University, Oxford, UK}
\author{H.~Panuganti}
\affiliation{CERN, Geneva, Switzerland}
\author{A.~Pardons}
\affiliation{CERN, Geneva, Switzerland}
\author{A.~Perera}
\affiliation{Cockcroft Institute, Daresbury, UK}
\affiliation{University of Liverpool, Liverpool, UK}
\author{J.~Pucek}
\affiliation{Max Planck Institute for Physics, Munich, Germany}
\author{A.~Pukhov}
\affiliation{Heinrich-Heine-Universit{\"a}t D{\"u}sseldorf, D{\"u}sseldorf, Germany}
\author{B.~R\'{a}czkevi}
\affiliation{Wigner Research Center for Physics, Budapest, Hungary}
\author{R.L.~Ramjiawan}
\affiliation{CERN, Geneva, Switzerland}
\affiliation{John Adams Institute, Oxford University, Oxford, UK}
\author{S.~Rey}
\affiliation{CERN, Geneva, Switzerland}
\author{O.~Schmitz}
\affiliation{University of Wisconsin, Madison, Wisconsin, USA}
\author{E.~Senes}
\affiliation{CERN, Geneva, Switzerland}
\author{L.O.~Silva}
\affiliation{GoLP/Instituto de Plasmas e Fus\~{a}o Nuclear, Instituto Superior T\'{e}cnico, Universidade de Lisboa, Lisbon, Portugal}
\author{C.~Stollberg}
\affiliation{Ecole Polytechnique Federale de Lausanne (EPFL), Swiss Plasma Center (SPC), Lausanne, Switzerland}
\author{A.~Sublet}
\affiliation{CERN, Geneva, Switzerland}
\author{A.~Topaloudis}
\affiliation{CERN, Geneva, Switzerland}
\author{N.~Torrado}
\affiliation{GoLP/Instituto de Plasmas e Fus\~{a}o Nuclear, Instituto Superior T\'{e}cnico, Universidade de Lisboa, Lisbon, Portugal}
\author{P.V.~Tuev}
\affiliation{Budker Institute of Nuclear Physics SB RAS, Novosibirsk, Russia}
\affiliation{Novosibirsk State University, Novosibirsk, Russia}
\author{M.~Turner}
\affiliation{CERN, Geneva, Switzerland}
\affiliation{LBNL, Berkeley, CA, USA}
\author{F.~Velotti}
\affiliation{CERN, Geneva, Switzerland}
\author{L.~Verra}
\affiliation{Max Planck Institute for Physics, Munich, Germany}
\affiliation{CERN, Geneva, Switzerland}
\affiliation{Technical University Munich, Munich, Germany}
\author{J.~Vieira}
\affiliation{GoLP/Instituto de Plasmas e Fus\~{a}o Nuclear, Instituto Superior T\'{e}cnico, Universidade de Lisboa, Lisbon, Portugal}
\author{H.~Vincke}
\affiliation{CERN, Geneva, Switzerland}
\author{C.P.~Welsch}
\affiliation{Cockcroft Institute, Daresbury, UK}
\affiliation{University of Liverpool, Liverpool, UK}
\author{M.~Wendt}
\affiliation{CERN, Geneva, Switzerland}
\author{M.~Wing}
\affiliation{UCL, London, UK}
\author{J.~Wolfenden}
\affiliation{Cockcroft Institute, Daresbury, UK}
\affiliation{University of Liverpool, Liverpool, UK}
\author{B.~Woolley}
\affiliation{CERN, Geneva, Switzerland}
\author{G.~Xia}
\affiliation{Cockcroft Institute, Daresbury, UK}
\affiliation{University of Manchester, Manchester, UK}
\author{M.~Zepp}
\affiliation{University of Wisconsin, Madison, Wisconsin, USA}
\author{G.~Zevi~Della~Porta}
\affiliation{CERN, Geneva, Switzerland}
\collaboration{The AWAKE Collaboration}

\noaffiliation

\date{\today}

\begin{abstract}
We present numerical simulations and experimental results of the self-modulation of a long proton bunch in a plasma with linear density gradients along the beam path. %
Simulation results agree with the experimental results reported in~\cite{gradexp}: with negative gradients, the charge of the modulated bunch is lower than with positive gradients.
In addition, the bunch modulation frequency varies with gradient. %
Simulation results show that dephasing of the wakefields with respect to the relativistic protons along the plasma is the main cause for the loss of charge. %
The study of the modulation frequency reveals details about the evolution of the self-modulation process along the plasma. %
In particular for negative gradients, the modulation frequency across time-resolved images of the bunch indicates the position along the plasma where protons leave the wakefields. %
Simulations and experimental results are in excellent agreement. 

\end{abstract}


\maketitle

\section{Introduction \label{sec:intro}}

A plasma wakefield accelerator (PWFA)~\cite{PWFAChen} uses a relativistic particle bunch as driver. %
The energy gain by a witness bunch (electrons, positrons, muons, etc.) is smaller than, or equal to the energy lost by the drive bunch. %
Two options exist to produce very high-energy (TeV) witness bunches carrying kJ of energy, i.e., with $\approx 10^{10}$\,particles: staging of multiple plasma sections with wakefields driven by $\approx 100$\,J energy level bunches~\cite{staging} or using a single plasma 
with wakefields driven by a multi-kJ energy bunch, such as a proton bunch~\cite{SMI}. %
High-energy proton bunches available today have a root mean square (rms) length $\sigma_z$ of 6 to 12\,cm (corresponding to an rms duration $\sigma_t$ of 200 to 400\,ps), e.g., CERN SPS and LHC, and Brookhaven RHIC. %

In a plasma with density $n_{e0}$, a bunch with a Gaussian longitudinal profile can effectively drive wakefields when the plasma wavelength $\lpe$ is on the order of its rms length $\sigma_z$ as $\lpe = \left(\frac{\pi}{n_{e0}r_e}\right)^{1/2} \approx \sigma_z$, where $r_e$ is the classical electron radius.
In that case, the amplitude of the driven accelerating wakefields is on the order of the wave-breaking field~\cite{Dawson} $E_{WB}(\sigma_z) = 2\pi\frac{m_e c^2}{e}\frac{1}{\sigma_z}$, where $m_e$ is the rest electron mass, $e$ its charge, and $c$ is the speed of light in vacuum. %
This scaling yields $< 55$\,MV/m fields for these long proton bunches. %
To drive wakefields with an amplitude in the order of GV/m, a long proton bunch must undergo the process of self-modulation (SM)~\cite{SMI}.  

Self-modulation occurs when a long, relativistic, charged particle bunch propagates in a dense plasma, i.e., when $\sigma_z \gg \lpe$. %
To avoid filamentation~\cite{bib:filamentation} the bunch must have an rms width $\sigma_r < \lpe$. %
The SM transforms the long bunch into a train of microbunches shorter than, and with a periodicity of approximately $\lpe$. %
The bunch train then resonantly drives wakefields with amplitudes that can reach a significant fraction of $E_{WB}(\lpe)$. 
The plasma density can thus be adjusted so the amplitude of the wakefields can exceed the GV/m level in a plasma with $n_{e0} > 10^{14}\,\mr{cm}^{-3}$, even with a cm-long bunch. %

During the SM growth of a long beam, the phase velocity of the wakefields is slower than the velocity of the drive bunch $v_b$ and evolves along the bunch $\xi = z - ct$ and plasma $z = ct$~\cite{phasePukhov,phaseSchroeder,gradexp}. %
The difference between the phase velocity and the bunch velocity is given by~\cite{phasePukhov} %

\begin{equation}
|\Delta v| \approx \frac{1}{2}\left(\frac{\xi}{z}\right)^{1/3}\left(\frac{n_b m_e}{n_{e0}m_p\gamma_p}\right)^{1/3}v_b,
\label{eqn:vpukhov}
\end{equation}

where $n_b$ is the bunch density, $m_p$ is the proton mass, and $\gamma_p$ is the gamma factor of the protons. This slow velocity causes a phase shift between the microbunches and the wakefields during the SM growth and possibly after that. %
The effect of $\Delta v$ is cumulative along the bunch and thus more important at the back of the bunch than in its front. %
This means that particles of the drive bunch do not necessarily remain in the focusing or defocusing phase of the wakefields all along their propagation in the plasma, which affects the process of the microbunch train formation. %
Similarly, they do not necessarily remain in the accelerating or decelerating phase of the wakefields, which also affects the amplitude of the wakefields.   %

The use of a positive plasma density gradient was proposed to produce an effective positive additional phase velocity to compensate at least for some of the deleterious effects of phase velocity variations during SM growth~\cite{phasePukhov,phaseSchroeder}. %
The effect of positive gradients is to shorten $\lpe$ along the bunch path, causing any given constant phase point to move toward the seeding position (Eq.~\ref{eqn:vpukhov}), thereby counteracting the backwards phase shift. %
The effect of negative gradients is to lengthen $\lpe$, enhancing the phase shift.
This is most important for the acceleration of electrons injected early along the plasma. %
In a constant density plasma, i.e., no gradient, these electrons could quickly dephase in the slow wakefields and gain little energy or even be defocused and lost. %
Plasma density gradients are also useful for maintaining particles of the drive bunch in phase with the wakefields, leading to trains with more microbunches~\cite{gradexp}, possibly leading to larger accelerating fields~\cite{Gorn_2020}. %

We showed the effects of plasma density gradients $g$ on the SM of a 400\,GeV proton bunch in an experimental paper~\cite{gradexp}. %
These results clearly show that positive gradients along the plasma lead to a higher number of microbunches in the train with more charge per microbunch. %
Measurements also show that the bunch modulation frequency $\fmod$, measured after 10\,m of plasma, changes as a function of gradient value.

While the experimental results show many of the features expected in the presence of density gradients, limitations with access to the SM process (observation of the bunch only after propagation in plasma and vacuum), as well as the lack of plasma density perturbation diagnostics, restrict the ability to measure many characteristics of the SM process along the plasma and as a function of $g$ values. %
However, in numerical simulations one has access to the beam, plasma, and wakefields characteristics all along the plasma. %

Here we show simulation results obtained with the particle-in-cell code OSIRIS 4.4.4~\cite{osiris} using parameters similar to those of the experiment. %
We show that there is a remarkable agreement between simulation and experimental results. 
We look into the details of the evolution of the proton bunch and wakefields along the plasma to explain the results obtained after the plasma. %
We relate the amount of charge in the core of the microbunch train to the dephasing of the wakefields with respect to the proton bunch, as well as to the amplitude of the wakefields. %
We show that the off-axis, time-resolved proton distribution carries information about $\fmod$ earlier along the plasma and that $\fmod$ for the microbunch train core evolves very differently along $z$ for positive and negative gradients. %
We also show that, among the values we use, there is indeed a positive gradient value that leads to a larger amplitude of the transverse wakefields when compared to the constant density case. %
We show however that a linear density gradient does not eliminate the issue of decreasing amplitude of the wakefields along the plasma past the SM saturation point. %

\section{General Setup \label{sec:setup}}

The SPS at CERN provides the proton bunch for the experiment. %
For the experimental results presented here, it has a population of $(2.98 \pm 0.16)\times 10^{11}$ protons, each with an energy of 400\,GeV. %
The rms duration of the bunch is $\sigma_t = 230$\,ps (or equivalently a rms length $\sigma_z = 6.9$\,cm). %
As we use the same data as was used in~\cite{Gorn_2020}, we also consider that the bunch is focused to an rms transverse size of $\sigma_{r0} \approx 200$\,\textmu m at the entrance of the 10\,m-long plasma and that it has a normalized emittance $\epsilon_N = 3.6$\,mm-mrad. %

The SM process is seeded, in the experiment by a relativistic ionization front (RIF)~\cite{gordon,modulationexp,SMISSM}, and in simulations by a step in the density profile of the proton bunch at the same position along the bunch as that of the RIF. %
Here, the two seeding methods are considered to be equivalent. %

The plasma has a linear density gradient $g$ in a $\pm$2\,\%/m range. 
With $g = 0$\,\%/m the density is constant along the plasma. %
In the experiment, the measured $g$ values are: $-1.93, -0.93, -0.52, +0.03, +0.43, +0.87, +1.30,$ and $+2.00$\,\%/m. %
In simulations, we use nine values: $-2, -1.5, -1, -0.5, 0, +0.5, +1, +1.5,$ and $+2\,$\%/m, in order to cover the same range as that of the experiment. %
In the rest of the text and for simplicity, we refer to experimental measurement using their respective closest half-integer value, e.g., $g = -2$\,\%/m for $g = -1.93$\,\%/m. %

\section{Simulation parameters \label{sec:simpar}}

\renewcommand{\arraystretch}{0.5} 
\begin{footnotesize}
\setlength{\tabcolsep}{12pt}
\begin{table*}[htbp]
  \centering
  \caption{Simulation parameters, where $\omega_{pe} = \sqrt{4\pi c^2 n_{e0} r_e}$.}
    \begin{tabular}{lcc}
    \textbf{Plasma and window param.} & \textbf{Phys. value} & {\textbf{Norm. value}} \\
    \midrule
    Initial plasma density & $1.81 \times 10^{14} \mr{cm}^{-3}$ & 1 $n_{e0}$ \\
    Plasma radius & 0.15 cm & 3.8 $c/\omega_{pe}$ \\
    Plasma length & 10.2 m & 25850 $c/\omega_{pe}$ \\
    Simulation window length & 21 cm & 532 $c/\omega_{pe}$ \\
    Simulation window width & 0.158 cm & 4 $c/\omega_{pe}$ \\
    Longitudinal resolution & 5.9 \um & 0.015 $c/\omega_{pe}$ \\
    Transverse resolution & 4 \um & 0.01 $c/\omega_{pe}$ \\
    Time step & 9.2 fs & 0.007  $\omega_{pe}^{-1}$ \\
    Particles per cell & --    & $3 \times 3$ \\
          &       &  \\
    \textbf{Bunch param.} &       &  \\
    \midrule
    RMS radius ($\sigma_{r0}$) & 200 \um & 0.51  $c/\omega_{pe}$ \\
    RMS length ($\sigma_z$) & 6.9 cm & 176.9  $c/\omega_{pe}$ \\
    Norm. emit. ($\epsilon_N$) & 3.6 mm mrad & 0.018 $m_p c$ \\ 
    Seed position (ahead of bunch center) & 3.81 cm & 96.4 $\omega_{pe}^{-1}$ \\
    Relativistic factor ($\gamma_p$) & 426.44 & 426.44 \\
    Relative energy spread & 0.035 \% & 0.035 \% \\
    Population & $3\times 10^{11}$ protons & -- \\
    Peak density & $6.9 \times 10^{12}\,\mr{cm}^{-3}$ & 0.038 $n_{e0}$ \\
    Particles per cell & --    & $2 \times 2$ \\
    \end{tabular}%
  \label{tab:simpar}%
\end{table*}%

\end{footnotesize}

We perform simulations in 2D axisymmetric geometry, using the parameters in Table \ref{tab:simpar} chosen after suitable convergence tests. %
We also compared results with a wider simulation window (0.32 cm) for selected $g$ values, and showed that for this study the halo effects~\cite{Gorn_2020} are negligible and we can use a narrow window to save computation time. %
The use of 2D simulations excludes 3D effects, which could have appeared in the experiment, such as the hosing instability~\cite{bib:HosingSchroeder}. %
However, 2D simulations reproduce the SM and many of the effects observed experimentally, while no severe hosing was observed. %
The simulation window is moving at $c$ together with the proton bunch. %
The simulation output data is saved every $\approx 10\,$cm (i.e., every 36928 time steps). %
We propagate simulation particles in vacuum to the locations of the screens in the experiment, 2\,m and 3.5\,m downstream from the plasma end. %

In the experiment, bunch %
parameters are measured as a function of time at a fixed position (that of the second screen). %
In the simulations, all quantities are measured at a fixed time, over the spatial extent of the proton bunch. %
However, the evolution of the proton distribution over its duration or length is sufficiently small that the results do not have significant differences. %

\section{Experimental parameters and diagnostics \label{sec:exppar}}

The plasma is created by a 120\,fs-long laser pulse, with a radius of $\approx 1\,$mm and an energy of $\approx$ 110\,mJ, ionizing a Rb vapor and co-propagating within the proton bunch, 128\,ps ahead of its density peak. %
The laser pulse ionizes the outermost electron of each Rb atom. %
It creates a RIF much shorter than the period of the wakefields ($> 8$\,ps). %
The interaction of the bunch and the plasma starts suddenly at the RIF, which provides seed wakefields for the SM to grow from~\cite{modulationexp,SMISSM}. %
When seeded, the phase of the bunch modulation with respect to the RIF is reproducible, from event to event, within a small fraction of a modulation period~\cite{SMISSM}. %

The plasma density is obtained by measuring the Rb vapor density with white-light interferometry with an uncertainty of 0.5\,\%~\cite{interferometry}. %
The Rb and plasma densities agree with each other~\cite{modulationexp}. %
We therefore quote the plasma density hereafter, even though the Rb density is measured. %

The density gradient is created by controlling the temperature of the reservoirs that evaporate Rb into the source at the plasma entrance and exit~\cite{gennady,vaporsource}. %
The experimental density gradient is calculated by dividing the difference in densities at the entrance and exit of the plasma by its length. %
In all cases, we keep the density at the plasma entrance, $n_{e0}=1.81 \times 10^{14}$\,cm$^{-3}$, constant and vary that at the exit. %

After the plasma, the proton bunch propagates in vacuum towards screens where it is characterized. %
We image onto the entrance slit of a streak camera the backward, incoherent, optical transition radiation (OTR) produced when protons enter the metallic screen located 3.5\,m downstream from the plasma exit~\cite{modulationexp,streak}. %
The camera gives a time-resolved image of the transverse density distribution of the proton bunch in a $\approx 74$\,\textmu m-wide slice of the bunch about its axis. %
We obtain experimental images from multiple short-time-scale (209\,ps) images stitched together in time~\cite{FabianStitching}. %
The fact that these stitched images form long trains of microbunches (see Fig.~\ref{fig:allprofiles} (a,c,e)) as opposed to a blur, confirms that the SM process is in the seeded~\cite{SMISSM} and not in the instability regime~\cite{SMI}. %

\section{Density profiles of the microbunch train \label{sec:density}}

\begin{figure*}[!ht]
    \centering
    \includegraphics[width=12.9cm]{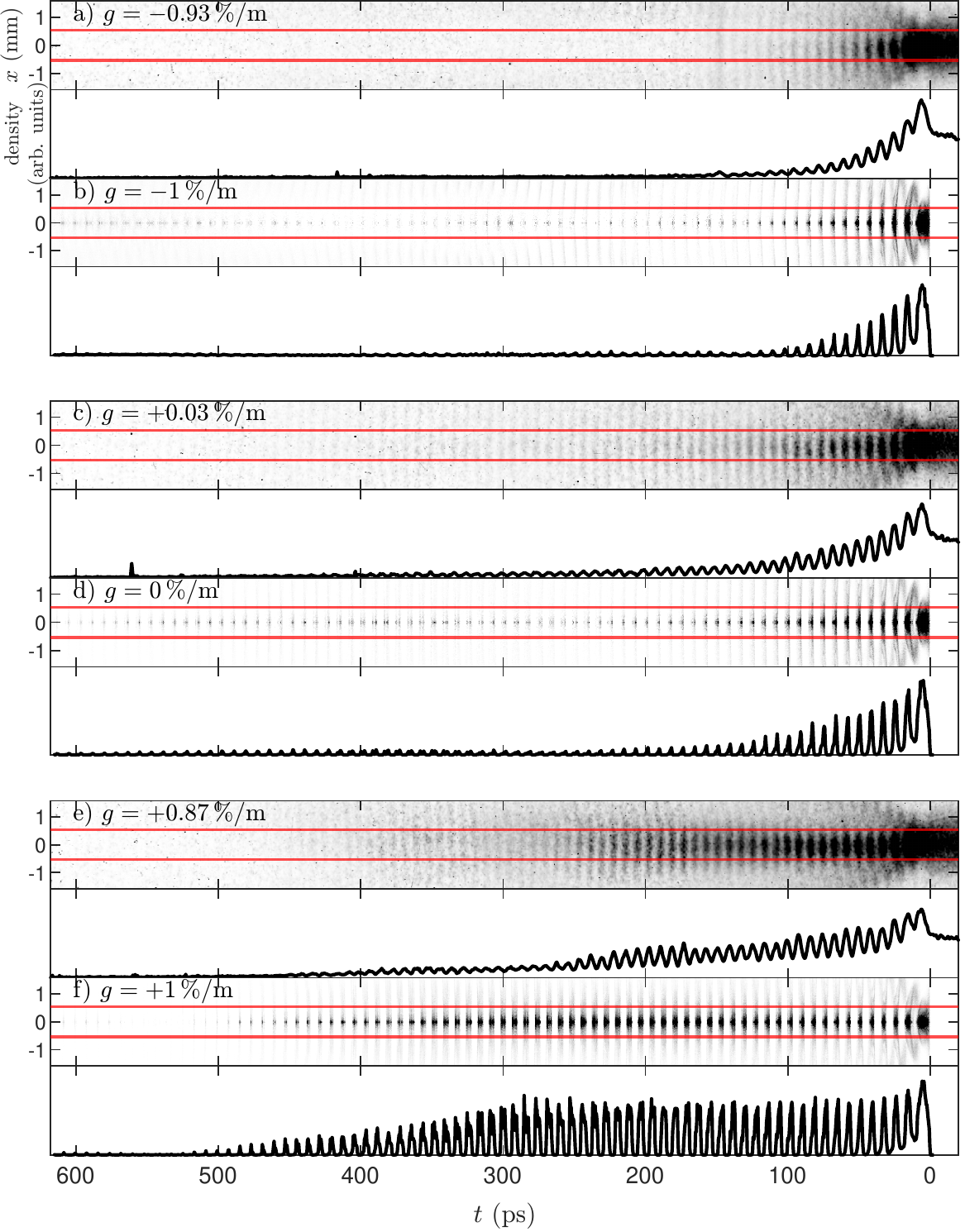} 
    \caption{Time-resolved experimental (a,c,e) and simulation (b,d,f) images and profiles of the modulated bunch with $g = -1$\,\%/m (a,b),  $g = 0$\,\%/m (c,d), and  $g = +1$\,\%/m (e,f). %
    Longitudinal profiles obtained by summing counts within $\sigma_{r,\mr{screen}}$: $|x|=\pm 0.536$\,mm (red lines on image) of the axis. %
    Images from 2D simulations are mirrored about the bunch axis for a more direct comparison with experimental ones. \label{fig:allprofiles}}
\end{figure*}

We display in Fig.~\ref{fig:allprofiles} time-resolved images of the charge density distribution from the experiment~\cite{streak} and from simulations for selected $g$ values, together with their on-axis time profiles. %
These images are part of a series that includes all measured $g$ values. %
The experimental images for $g = -2$ and $+2$\,\%/m are displayed in~\cite{gradexp} and together with the images displayed here show the same features. %
We emphasize here the similarities and differences between experimental and simulation images. %

We generate the experimental profiles by summing counts of the images up to a transverse extent of $|x|= \sigma_{r,\mr{screen}} = 0.536$\,mm from the axis (red lines on Fig.~\ref{fig:allprofiles}), where $\sigma_{r,\mr{screen}}$ is the rms width of the incoming (no plasma) bunch at the screen position. %
We obtain this width from a Gaussian fit to the transverse profile of the part of the bunch ahead of the ionization front ($t < 0$\,ps on Fig.~\ref{fig:allprofiles} (a,c,e)). %
In simulations, we sum the charge of macroparticles in bins with size equal to that of a pixel from the experimental images, and divide by the volume in the ring corresponding to each bin (2D simulations). %
We calculate on-axis time profiles by summing the density profiles transversely up to $\sigma_{r,\mr{screen}}$, as for the experimental images. %

It is clear that the resolution of the simulation images is better than that of the experimental ones. %
These are subject to the space ($\approx 180$\,\textmu m) and time resolution ($\approx 3$\,ps) of the diagnostic. %
We use $g = 0$\,\%/m as reference to compare to the other two cases. %

Figure~\ref{fig:allprofiles} shows that images (c) from the experiment and (d) from simulations with $g = 0$\,\%/m look quite similar. 
In both cases, the density of the microbunch train decreases from the $t = 0$\,ps seeding position to $t \approx 200$\,ps. %
After that point, there is a series of microbunches with a small and relatively constant density when compared to that of the ones at the front. %
We also note that as a result of focusing, the first two or three microbunches have a charge density larger than that of the incoming bunch ($t < 0$\,ps on experimental images, not shown on simulation images). %

Images with negative gradient ($g = -1$\,\%/m, Fig.~\ref{fig:allprofiles} (a,b)) also show similar characteristics between simulations and experiment. %
In this case, the microbunch train is shorter than with $g = 0$\,\%/m. %
Only a few microbunches remain, between $t = 0$\,ps and $t \approx 100$\,ps. %
The defocused charge can be seen away from the axis, outside the red lines at $t > 100$\,ps. %

In the positive gradient case ($g = +1$\,\%/m, Fig.~\ref{fig:allprofiles} (e,f)), the charge density of the microbunches remains approximately constant up to $t \approx \sigma_t = 230$\,ps. %
In that region there are 27 microbunches at this plasma density, both in simulations and in the experiment. %
After that point, there is a sudden decrease of the charge density per microbunch on the experimental image and profile (Fig.~\ref{fig:allprofiles} (e)). %
In the simulation result (Fig.~\ref{fig:allprofiles} (f)), the charge density starts decreasing after $t \approx 300$\,ps and reaches a value close to zero at $t \approx 500$\,ps. %

This quick analysis of Fig.~\ref{fig:allprofiles} shows that there is a good general agreement between experimental and simulation results. %
It also shows that experimental results with possible 3D effects (Fig.~\ref{fig:allprofiles} (e)) differ somewhat from simulation results obtained in 2D (Fig.~\ref{fig:allprofiles} (f)), especially at later times along the bunch. %
Since the bunch train is generally longer with $g > 0$, the measurement is more sensitive to possible misalignment between the bunch and the streak camera slit. %
With the longer bunch train with more charge, the non-axisymmetric hose instability~\cite{bib:HosingSchroeder} is more likely to develop and steer the bunch centroid away from, and sideways along, the slit of the camera as shown by the slight wiggles of the microbunches around the axis (between $t \approx 100$\,ps and $\approx 250$\,ps) in Fig.~\ref{fig:allprofiles} (e).
These effects may explain the longer train observed in simulations with $g = +1$\,\%/m. %
Figure~\ref{fig:allprofiles} shows that the effect of the plasma density gradient on the bunch charge density distribution can be threefold: %
change the length of the bunch train, i.e., the number of microbunches in the train; %
change the charge density and charge of individual microbunches; %
increase the likelihood of developing observable non-axisymmetric effects. %

\section{Charge of the modulated proton bunch \label{sec:charge}}

\begin{figure}[h]
    \centering
    \includegraphics[width=8.6cm]{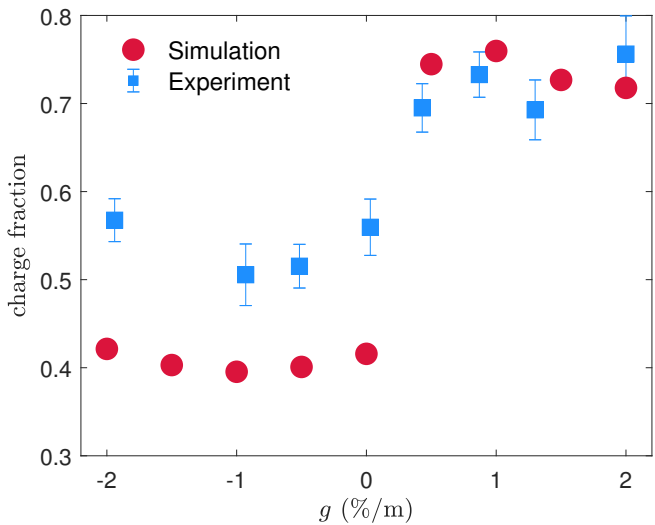}
    \caption{Proton bunch charge fraction within one rms width at the screen of the unmodulated proton bunch versus $g$ from experimental~\cite{gradexp} (blue squares) and from simulation (orange dots) results. %
    The number of measurements considered for the average of the experimental data are, from $g = -2$ to $+2$\,\%/m: 60, 51, 33, 38, 10, 14, 16, and 29. %
    Error bars are standard deviations for each data set. \label{fig:bunchfrac}}
\end{figure}

In the experiment, we determined the charge fraction in the core of the microbunch train (charge within one rms width normalized to the charge in the same volume of an unmodulated proton bunch) from time-integrated images of the transverse bunch distribution for each $g$ value. %
We obtain these images 2\,m downstream from the plasma exit. %
The results were published in~\cite{gradexp} and are reproduced in Fig.~\ref{fig:bunchfrac} (blue squares). %
The figure shows that there is a clear difference in the amount of charge measured with $g > 0$ and $g \le 0$. %
As mentioned in~\cite{gradexp}, charge variations can be attributed to changes in phase velocity and amplitude of the wakefields along the plasma due to the density gradient. %

With simulation data we obtain the charge of the microbunch train that is within one rms width of a bunch propagating in vacuum, after 2\,m of propagation from the plasma end. %
We then normalize it to the charge of an unmodulated bunch in the same volume. %
We finally add the fraction of charge ahead of the step in the density profile ($\approx 29$\,\%), which is not present in simulations, but is in the experiment. %

Figure \ref{fig:bunchfrac} shows that there are essentially two charge fraction values, one for $g > 0$ and one for $g \le 0$. %
The values are similar between simulation and experiment for $g > 0$, while the simulation values are $\approx 0.15$ lower than the experimental ones for $g \le 0$. %

\begin{figure}[ht]
    \centering
    \includegraphics[width=8.6cm]{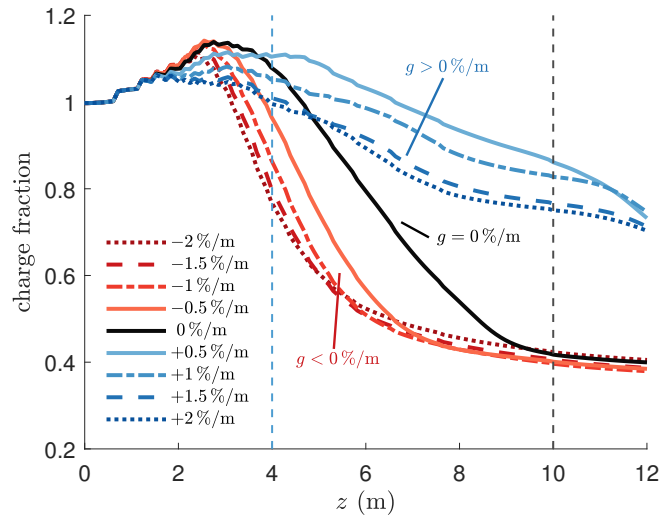}
    \caption{Proton bunch charge fraction within one $\sigma_r$ radius of the unmodulated proton bunch along $z$, for various $g$ values, calculated from simulation results. %
    Black dashed line: position of plasma end. %
    Blue dashed line: position of the peak in the mean defocusing wakefields for $g = 0$\,\%/m (see Fig.~\ref{fig:meandefocusing}).
    \label{fig:charge_evolution}}
\end{figure}

To better understand this feature, we now investigate with numerical simulations the evolution of the bunch charge fraction along the plasma, displayed in Fig.~\ref{fig:charge_evolution}. We calculate it as explained above, as a function of propagation distance in the plasma ($z < 10$\,m) and until the screen ($z \geq 10$\,m) for all $g$ values. %
We calculate the bunch transverse size at each $z$-position: $\sigma_{r}(z) = \sigma_{r0}\left(1+\frac{z^2\epsilon_N^2}{\gamma_p^2\sigma_{r0}^4}\right)^{1/2}$. %
The charge fraction values at $z = 12$\,m in Fig.~\ref{fig:charge_evolution} correspond to the ones on Fig.~\ref{fig:bunchfrac}. %

The figure shows that, in all cases, the charge fraction increases over the first 3\,m due to the overall focusing effect of the transverse wakefields and of the plasma adiabatic response. %
After that there is significant loss of charge due to the SM process and microbunch train formation. %
Figure~\ref{fig:charge_evolution} shows that past $z \approx 5$\,m, the charge fraction continues to decrease for all $g$ values due to the continuous evolution of the wakefields and proton distribution. %

\begin{figure}[h]
    \centering
    \includegraphics[width=8.6cm]{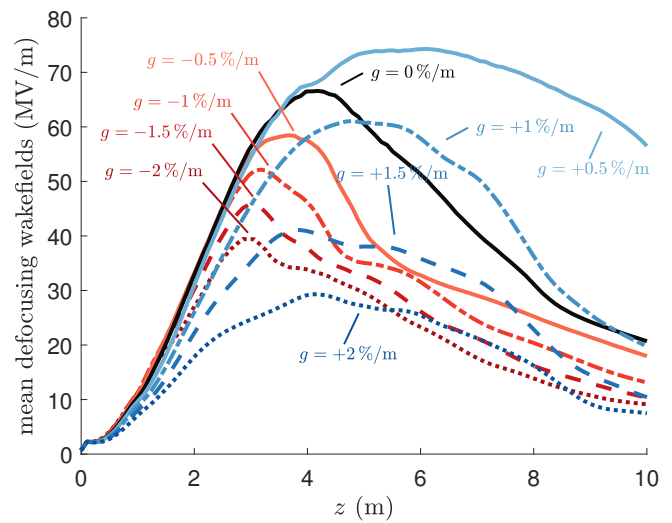}
    \caption{Mean amplitude of the defocusing wakefields, averaged over the whole simulation window, along $z$ for various $g$ values, from simulation results.  \label{fig:meandefocusing}}
\end{figure}

Figure~\ref{fig:meandefocusing} (similar to Fig. 9 in \cite{Gorn_2020}) shows that the mean defocusing wakefields~\footnote{We plot the mean value of the wakefields, averaged over the whole simulation window, because transverse wakefields depend on the position along the bunch ($\xi$) and across the bunch (radius or $x$) and protons may not be subject to their maximum value or to the value at the incoming bunch radius location (usually plotted).}
reach their maximum amplitude, i.e., saturation, between 3 and 5.5\,m, sooner for $g < 0$ (red lines) and later for $g > 0$ (blue lines). %
For $g < 0$, the charge fraction decreases more than for $g > 0$, and it does so faster immediately after saturation, between $z \approx 3$\,m and $z \approx 7$\,m, and slower afterwards. %
The charge fraction value for $g = 0$\,\%/m (black line) is between the ones for $g \neq 0$\,\%/m and ends with a value similar to those of $g < 0$ at the measurement location. %
We note here that Fig.~\ref{fig:meandefocusing} shows that $g = +0.5$\,\%/m yields the largest mean transverse wakefield amplitudes and it is the only one with a significantly larger value than $g = 0$\,\%/m. 

\begin{figure*}[!ht]
    \centering
    \includegraphics[width=17.2cm]{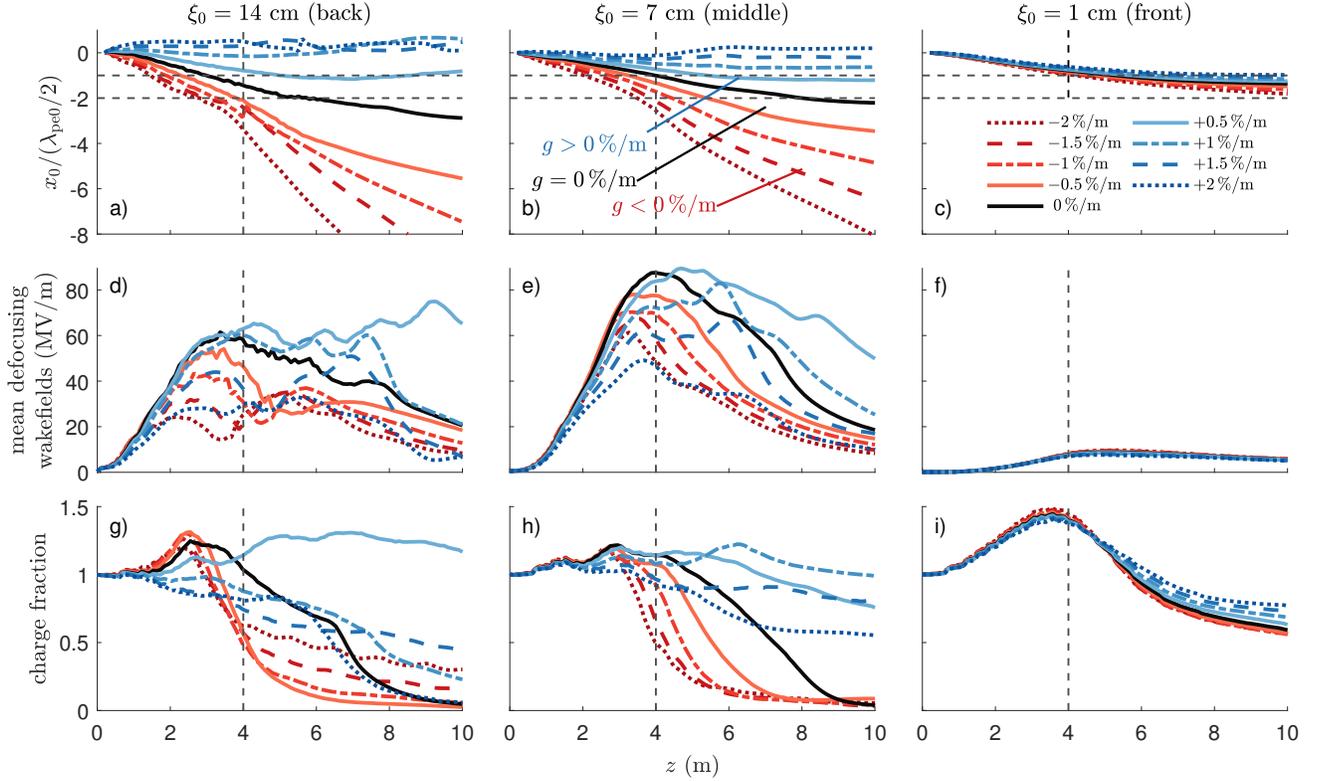}
    \caption{(a) to (c): Position $x_0$ of the zero-crossing of the longitudinal wakefields on axis, normalized to half the initial plasma wavelength $\lzero/2$. %
    (d) to (f): Mean amplitude of the defocusing wakefields within $\xi_0 \pm \lpe(z)/2$. %
    (f) to (i): Charge fraction within $\xi_0 \pm \lpe(z)/2$ and within one $\sigma_r(z)$. %
    All quantities plotted as a function of $z$ starting at three positions along the bunch: (a,d,g) $\xi_0 = 1$\,cm; (b,e,h) $\xi_0 = 7$\,cm; and (c,f,i) $\xi_0 = 14$\,cm. %
    Horizontal dashed lines: $-1(\lzero/2)$ and $-2(\lzero/2)$. %
    Vertical dashed lines: position of the peak in the mean defocusing wakefields for $g = 0$\,\%/m (see Fig.~\ref{fig:meandefocusing}). %
    Each set of curves in (g) to (i) is normalized to the charge within $\pm \lpe(z)/2$ and $\sigma_r(z)$ of the unmodulated proton bunch. The initial charge values ($z = 0$\,m) are: 235, 608, and 617\,pC respectively. %
    \label{fig:dephasing}}
\end{figure*}   

We investigate in simulations the evolution of the phase and amplitude of the wakefields to provide a more detailed explanation for the evolution of the charge fraction seen on Fig.~\ref{fig:charge_evolution}. %
Since the transverse position of the maximum value of transverse wakefields changes along the plasma, we use the longitudinal wakefields to analyze the phase of the wakefields. 
To get a handle on the phase, we follow the position $x_0$ of the zero-crossing of the on-axis longitudinal wakefields and its consequences on the local bunch charge along the plasma, as opposed to the evolution at fixed $\xi_0$ positions. %
The rationale is that since wakefields and charge distributions shift in phase along the plasma, we follow the wakefields rather than single protons. %
For that purpose we choose three representative initial positions along the bunch, near the step in the density profile in the front of the bunch ($\xi_{0} \approx 1$\,cm), after the peak of the bunch density and of the amplitude of the wakefields ($\xi_{0} \approx 7$\,cm), and to the rear of the bunch ($\xi_{0} \approx 14$\,cm). %

To calculate the charge fraction evolution and mean amplitude of the defocusing wakefields at the different $\xi_0$ positions, we consider that, with a linear plasma density gradient, the plasma wavelength changes along the plasma as $\lpe(z) = \lzero \left(1+\frac{g}{100}z\right)^{-1/2}$, with $\lzero = \lpe(z = 0\,\mathrm{m})$. %
Figure~\ref{fig:dephasing} (a-c) shows that the position of $x_0$ shifts backwards for $g = 0$\,\%/m and shifts more the further along the bunch. %
They also show that it does not shift according to a local adjustment of the plasma wavelength as $\lpe \propto n_e(z)^{-1/2}$, because it is a combination of the velocity difference described by Eq.~\ref{eqn:vpukhov}, the further evolution of the microbunch train after saturation, and the effect of the density gradient. The magnitude of the shift in $x_0$ is also different between $g > 0$ and $g < 0$, being larger in the latter. %

Early along the bunch ($\xi_{0} = 1$\,cm, $\approx 4\lzero$),  Fig.~\ref{fig:dephasing} (c) shows that the dephasing of the wakefields is small ($< 2(\lzero/2)$, wakefields change from focusing to defocusing or vice versa over $\lpe/2$) and their amplitude is small ($< 10$\,MV/m, Fig.~\ref{fig:dephasing} (f)) in all cases. %
Thus, most of the charge loss (Fig.~\ref{fig:dephasing} (i)) occurs after $z \approx 4$\,m, after the initial focusing phase, and is dominated by the continuous evolution of the wakefields past their saturation point. 
This early along the bunch, all quantities are similar among the various $g$ values because the cumulative effect of the change in period of the wakefields caused by the change in plasma density is small after only $\approx 4\lzero)$. %
The charge fraction remaining at the end of the plasma ($\approx 70\,\%$) is in general larger than further along the bunch (Fig.~\ref{fig:dephasing} (h,g)), because of the interplay between the low transverse wakefields protons are subject to, which allows them to be recaptured by the focusing wakefields after having been weakly radially pushed out by the defocusing ones. %

The situation is quite different near the middle of the bunch ($\xi_{0} = 7$\,cm, $\approx 28\lzero$). 
There, the gradients cause a split (Fig.~\ref{fig:dephasing} (b,h)) between $g > 0$ for which dephasing remains small ($\lesssim \lzero/2$) and $g \le 0$ for which dephasing exceeds $-\lzero/2$ (Fig.~\ref{fig:dephasing} (b)) and reaches $-8\lzero/2$ (for $g = -2$\,\%/m). %
The maximum amplitude of the mean defocusing wakefields exceeds 50\,MV/m in all cases and varies by a factor of almost two with $g$ (Fig.~\ref{fig:dephasing} (e)). %
However, with $g = +0.5, +1,$ and $+1.5$\,\%/m, despite 
the larger amplitude of the wakefields than earlier along the bunch (Fig.~\ref{fig:dephasing} (f)), the small dephasing leads to less charge loss. %
With these $g$ values one observes an effect close to what is desired from the SM process that may be obtainable with a plasma density step~\cite{lotovdensitystep}: small or no charge loss after saturation, signifying the driving of wakefields with constant amplitude over a long distance. %
That amplitude depends on the charge in each microbunch, but also on the position of the microbunches within the accelerating and decelerating wakefields. %
Thus, even with a similar amount of charge, amplitudes might differ. %
With $g > 0$, the charge fraction remains larger than 50\,\% at $z = 10$\,m, with the most charge for $g = +1$\,\%/m. %
Cases with $g \le 0$ have very low charge fraction values at $z = 10$\,m. %

At the back of the bunch ($\xi_{0} = 14$\,cm, $\approx 56 \lzero$), Fig.~\ref{fig:dephasing} (a) shows that the dephasing is similar to that observed in the middle of the bunch (Fig.~\ref{fig:dephasing} (b)), although cases with $g < 0$ have a larger dephasing. %
With $g > 0$, $x_0$ takes positive values, i.e., the wakefields become superluminal. %
The amplitudes of the wakefields (Fig.~\ref{fig:dephasing} (d)) are in general smaller than those in $\xi_{0} = 7$\,cm. %
With $g = +0.5$\,\%/m (Fig.~\ref{fig:dephasing} (g)), the charge fraction remains high at $z = 10$\,m and the phase (with respect to the bunch) after saturation remains relatively constant when compared to other $g$ values. %
This is the result of the focusing of the charge in the region $r > \sigma_r$ to the axis, in addition to considering off-axis charge trapped between focusing and defocusing wakefields, but still within $r < \sigma_r$, which leads to charge fraction values larger than 1. 
We note here that the charge at this $\xi_0 = 14$\,cm position along the bunch (235\,pC initially in the $\xi_0 \pm \lzero/2$ interval, $> 600\,$pC at the other two locations) weakly contributes to the total charge variations. %

These results are consistent with what is observed on the bunch images of Fig.~\ref{fig:allprofiles}. %
Simulation results show that the charge preservation or loss depends on a combination of the effects of the phase and amplitude of the wakefields at the location of the protons along the bunch and all along the plasma. %
The case with $g = +0.5\,$\%/m shows quite a constant phase velocity along the bunch, and in general the largest amplitude of the wakefields and a high charge fraction remaining. %

\section{Proton bunch modulation frequency \label{sec:freq}}

In the experiment, we demonstrated that, after propagating in 10\,m of plasma, the bunch modulation frequency is equal to the plasma frequency $\fpe$ as $\fmod = \fpe = \left( \frac{c^2 n_{e0} r_e}{\pi} \right)^{1/2}$~\cite{modulationexp}. We used different values for the plasma density, covering a range of one order of magnitude. For each value, the plasma density was kept constant (no gradient). 
When using a plasma with a linear density gradient ($g \ne 0$\,\%/m), $\fmod$ lies between the plasma frequencies at the plasma entrance and exit~\cite{gradexp}. %
Two different measurement methods were used and they showed good agreement: the CTR diagnostic~\cite{CTR} and applying a discrete Fourier transform (DFT) frequency analysis to time profiles of time-resolved images of the bunch 
such as those of Fig.~\ref{fig:allprofiles}. %

Carefully observing bunch images on Fig.~\ref{fig:allprofiles} (a-f), one notices that the charge distribution of the microbunches curves away from the beam axis, showing a general C-shape for $g = -1$\,\%/m (Fig.~\ref{fig:allprofiles} (a,b)). %
The distribution appears quite straight for $g = 0$\,\%/m (Fig.~\ref{fig:allprofiles} (c,d)). %
These curvatures are displayed also for $g = -2$ and $+2$\,\%/m in the inset of Fig.~\ref{fig:fvsrgm20}.
Assuming a modulation that starts at the seeding position, such features could be an indication of $\fmod$ variations across the charge distribution at that location. %
Knowing that with density gradients the plasma frequency varies along the plasma, it may be expected that protons reaching a larger radial position at the screen have left the wakefields earlier along the plasma. %

We performed a DFT analysis on time profiles from simulation and experimental images after propagating 10\,m in plasma and 3.5\,m in vacuum. %
In both cases, we used a time window starting 12\,ps behind the seeding position to exclude from the analysis the first microbunch, which is usually longer than the following ones~\cite{bib:annamaria}. %
The window extends to 467\,ps ($\approx 2\sigma_t$) behind the seeding position. %
The width of the DFT frequency bin for this time window is 2.2\,GHz. %
By zero-padding the profiles, the width is reduced to 0.3\,GHz, which is similar to that obtained from the accuracy of the plasma density measurement~\cite{interferometry}. 
We take the frequency of the highest peak in the DFT power spectrum %
as $\fmod$. %

\begin{figure}[ht]
    \centering
    \includegraphics[width=8.6cm]{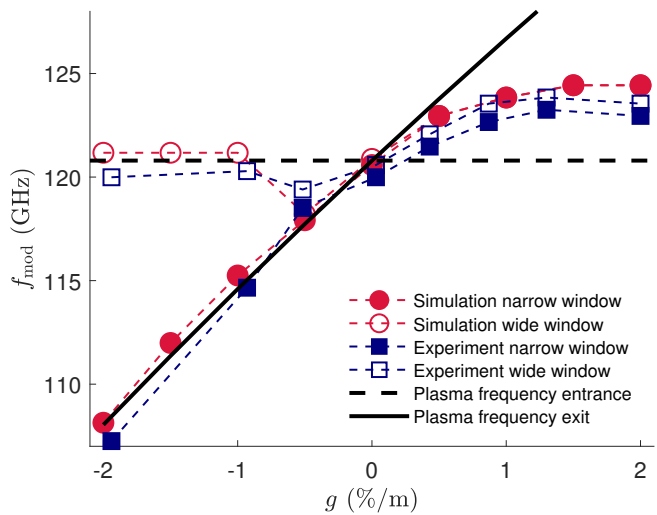}
    \caption{$\fmod$ as function of $g$, after propagating 10\,m in plasma and 3.5\,m in vacuum using a narrow (filled symbols, $|x| \le 0.36$\,mm) and a wide (empty symbols, $|x| = [0.36,1.8]$\,mm) transverse extent for the DFT analysis of time-resolved images, for both simulations (red symbols) and experiment (blue symbols). %
    Black lines: plasma frequencies at plasma entrance $\fzero$ (dashed) and exit $\fpe(g,z = 10\,\mr{m}) = \fzero\sqrt{1 + \frac{g}{100}z}$ (continuous). %
    }
    \label{fig:fvsg}
\end{figure}

Figure~\ref{fig:fvsg} shows $\fmod$ as measured in simulations (red symbols) and in the experiment (blue symbols), in both cases using a narrow transverse extent of $|x| \le 0.36$\,mm (filled symbols), that includes only the microbunch train, and a wide one of $|x| = [0.36,1.8]$\,mm (empty symbols) for the time profiles. %

Simulation and experimental results are in very good agreement with each other. %
The value of $\fmod$ is proportional to $\fpe(z = 10\,\mr{m})$ over the $-0.5\le g\le +0.5$\,\%/m range in simulations and experiments. 
For $g > +0.5$\,\%/m, $\fmod$ saturates at $\approx 3$\,\% above $\fzero$. %
For $g < -0.5$\,\%/m, $\fmod \approx \fzero$ in the wide window case, whereas it follows $\fpe(z = 10\,\mr{m})$ for the narrow window case. %

\begin{figure}[ht]
    \centering
    \includegraphics[width=8.6cm]{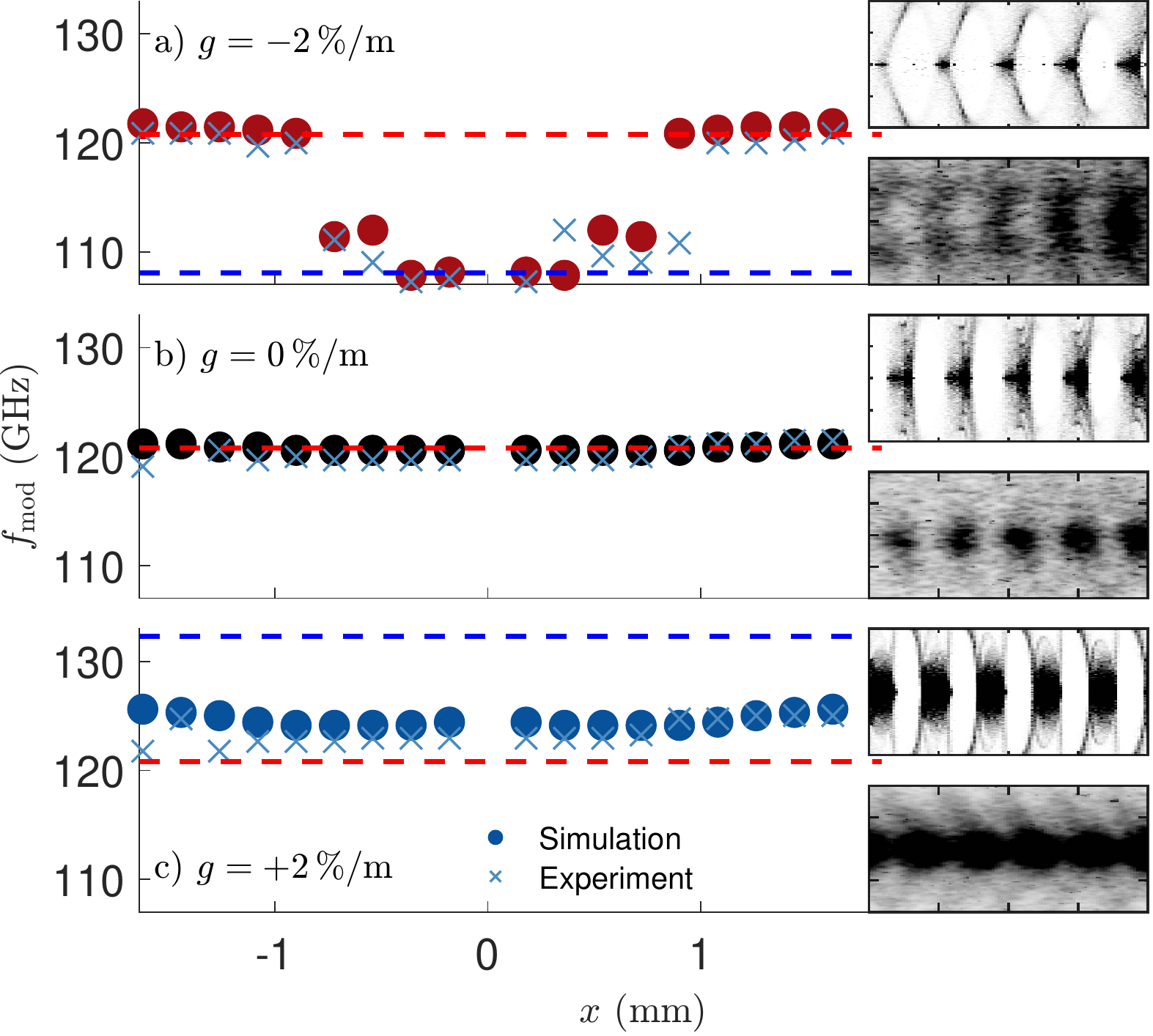}
    \caption{Modulation frequency $\fmod$ of 180\,\textmu m-wide slices along the transverse direction $x$ of the bunch from experimental (crosses) and simulation (circles) images with (a) $g = -2$\,\%/m, (b) $g = 0$\,\%/m, and (c) $g = +2$\,\%/m. %
    Simulation results (2D) are mirrored to compare to both sides of the experimental data. %
    Dashed red line: $\fzero$. %
    Dashed blue line: $\fpe(z = 10\,\mr{m})$. %
    Insets: time-resolved simulation (top) and experimental (bottom) images of the modulated bunch in a 40\,ps range (from the same series as those in Fig. \ref{fig:allprofiles}), showing the curvature of charge distributions for each $g$ value.}
    \label{fig:fvsrgm20}
\end{figure}

To further explore the behavior of $\fmod$ with respect to the distance from the axis, we perform a DFT analysis using multiple time profiles starting at the beam axis and with radial extents equivalent to the transverse resolution of the optical system in the experiment: 180\,\textmu m. %
For the figure, we mirror about the beam axis the frequency values from 2D simulations. %

Figure~\ref{fig:fvsrgm20} (a), for the gradient value that shows the largest difference in Fig.~\ref{fig:fvsg} ($g = -2$\,\%/m), shows that $\fmod$ is equal to $\fpe(z = 10\,\mr{m})$ near the axis ($|x| \le 0.36$\,mm) and transitions to $\fzero$ for $|x| \ge 0.96$\,mm. 
On the contrary, Fig.~\ref{fig:fvsrgm20} (b) shows that with $g = 0$\,\%/m the frequency remains close to $\fzero$ at all radii, in agreement with the single frequency expected. %
Figure~\ref{fig:fvsrgm20} (c) for $g = +2$\,\%/m, shows a slight increase in frequency with radius, unlike Fig.~\ref{fig:fvsrgm20} (a). %
There is very good agreement between the experimental and simulation results. %
The result for $g = -2$\,\%/m is consistent with the above hypothesis that protons reaching larger radii at the screen left the wakefields earlier along the plasma, carrying information about the plasma frequency at that location. %
In the other two cases (b,c), the variations are too small to draw strong conclusions from these results. 
The variation of $\fmod$ with radius also explains the curvature of the distributions visible on Fig.~\ref{fig:allprofiles} images and insets of Fig.~\ref{fig:fvsrgm20}. %

\begin{figure}[h]
    \centering
    \includegraphics[width=8.6cm]{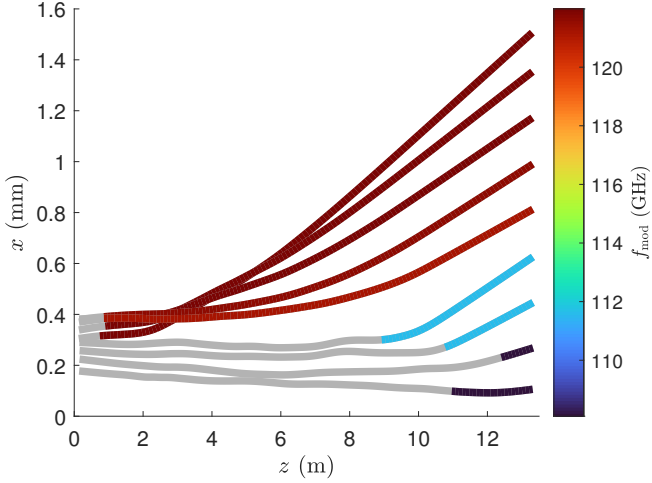}
    \caption{Mean radial position along the plasma and up to the screen position of simulation macroparticles that reach the screen position in 180\,\textmu m-wide slices for $g = -2$\,\%/m. %
    Lines are colored once the $\fmod$ of the distribution along the plasma matches the $\fmod$ measured at $z = 13.5$\,m. 
    }
    \label{fig:meantracksgm20}
\end{figure}

To confirm the origin of the proton distributions carrying certain frequencies, we back-track simulation particles along the plasma. %
We identify the macroparticles in each radial slice at the screen position ($z = 13.5$\,m) and calculate their mean radial position at each location along the plasma. %
Figure~\ref{fig:meantracksgm20} shows that indeed, for $g = -2$\,\%/m, particles reaching the larger radii at the screen left the wakefields early along the plasma ($z \approx 3$\,m) and carry a $\fmod$ equal to $\fzero$. %
The figure also shows in essence two populations: one that leaves the wakefields early and one that remains within the wakefields over most of the plasma length, both of them carrying the local plasma frequency as $\fmod$, as shown on Fig~\ref{fig:fvsrgm20} (a). %
The proton distributions, especially in the core, continue evolving even after the plasma, as they move towards or away from the axis due to their transverse momentum.

\begin{figure}[ht]
    \centering
    \includegraphics[width=8.6cm]{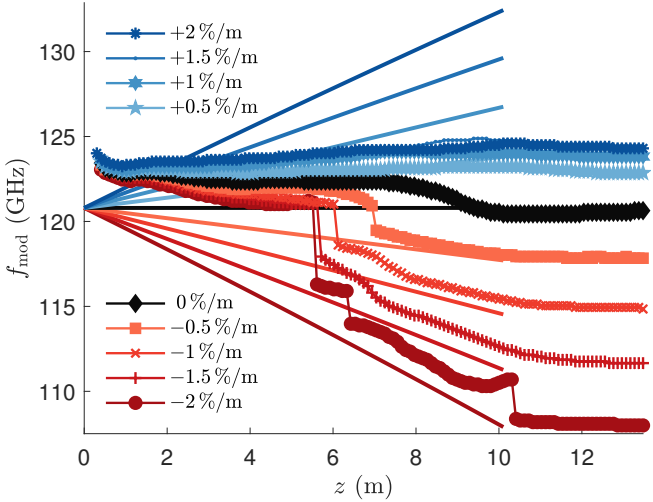}
    \caption{On-axis ($|x| \le 180$\,\textmu m) proton bunch modulation frequency $\fmod$ as a function of propagation distance for various $g$ values, from simulations. %
    Continuous lines: $\fpe(z)$ with corresponding $g$ value colors. %
    }
    \label{fig:fvsgz}
\end{figure}

We explore the difference in the frequency response among the various $g$ values by following the evolution of $\fmod$ near the bunch axis ($|x| < 180$\,\textmu m) along the plasma and up to the screen position. %

In Fig.~\ref{fig:fvsgz}, the modulation frequencies start around 124\,GHz, higher than $\fpe(n_{e0}) = 120.8$\,GHz, because the presence of the proton bunch density introduces an additional restoring force on plasma electrons, which increases their initial oscillation frequency: $\fpe(n_{e0} + n_{b}) = 123.3$\,GHz, and therefore also the initial $\fmod$. %
For all $g$ values, there is an initial decrease in $\fmod$ over the first meters of plasma caused by phase slippage during SM growth. 
Afterwards, as with the charge distribution and dephasing, there is again a different behavior between positive and negative gradient cases. %
The $\fmod$ for $g > 0$ is essentially constant along $z$, with relative changes between the minimum and maximum value of $\approx 1.5$\,\%. %
The bunch train is relatively long and thus cannot adjust well to the local plasma frequency, and its effectiveness at driving wakefields decreases for large $g$ values (see Fig.~\ref{fig:meandefocusing}), despite the large charge in the train and per microbunch (see Fig.~\ref{fig:bunchfrac}). %

It is interesting to note that the $\fmod = \fpe$ for $g = 0$\,\%/m reported in~\cite{modulationexp} is reached only near the end of the plasma, when the incoming bunch is fully modulated. %
The $g = 0$\,\%/m case has again an evolution somewhat between that of the $g > 0$ and $g < 0$ cases. %
The dominant $\fmod$ for $g < 0$ that is first slowly decreasing along the plasma, exhibits sudden jumps to basically follow $\fpe(z)$ after some point between $z = 5.5$\,m and $z = 7$\,m. %

\begin{figure}[ht]
    \centering
    \includegraphics[width=8.6cm]{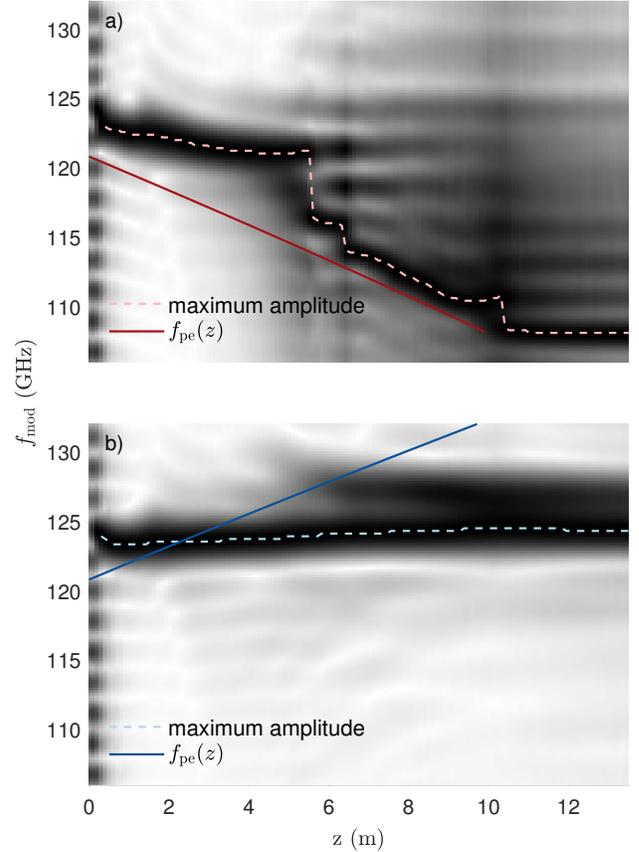}
    \caption{Waterfall plot of the modulated bunch DFT power spectrum along $z$ for (a) $g = -2$\,\%/m and (b) $g = +2$\,\%/m, from simulations. %
    Spectra normalized to their maximum at each $z$. %
    Continuous line: $\fpe(z)$, dashed line: frequency with maximum amplitude in each power spectrum (same line as on Fig.~\ref{fig:fvsgz}). %
    The periodic signal at $z \approx 0$\,m (and later) shows the artifact of the DFT caused by the finite time window duration of 455\,ps, corresponding to  $\Delta f \approx 2.2$\,GHz. %
    }
    \label{fig:fvszwaterfall}
\end{figure}

\begin{figure*}[!ht]
    \centering
    \includegraphics[width=15.2cm]{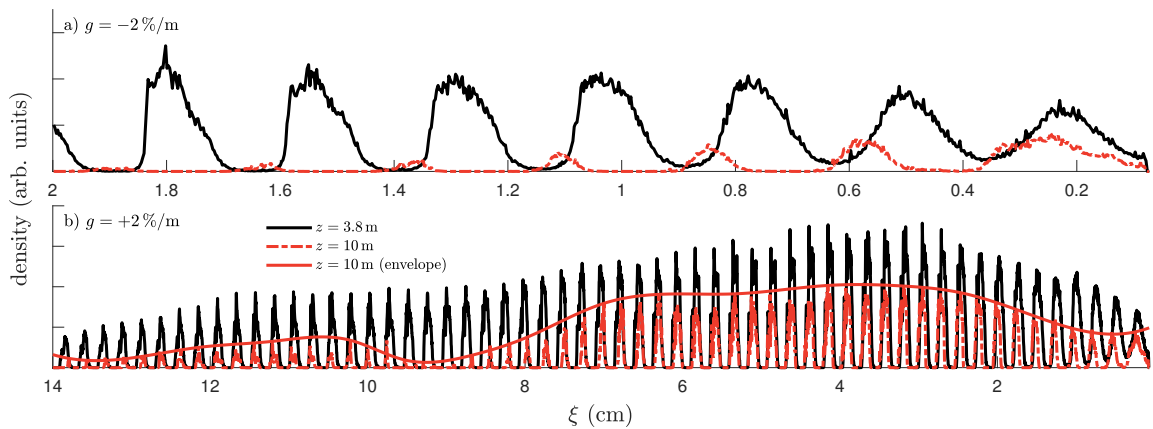} 
    \caption{Longitudinal profile of the microbunch train within $|x| < 180$\,\textmu m for (a) $g = -2$\,\%/m ($\xi < 2$\,cm, close to the bunch front) and (b) $g = +2$\,\%/m ($\xi < 14$\,cm) at two locations along propagation: close to saturation ($z = 3.8$\,m, black line), and plasma exit ($z = 10$\,m, red line), together with the envelope for $g = +2$\,\%/m (continuous red line). %
    }
    \label{fig:microbunch}
\end{figure*}

Figure~\ref{fig:fvsgz} shows limited information about the DFT power spectrum because it only displays the frequency with the highest amplitude. %
However, Fig.~\ref{fig:fvszwaterfall} shows that for (a) $g = -2$ and (b) $g = +2$\,\%/m  the DFT power spectrum does not have a single peak along $z$, and amplitude variations in the wakefields and bunch density modulation, as well as structures in the transverse charge distributions caused by successive defocusing and focusing of protons create multiple spectral features. %
For $g = -2$\,\%/m (Fig.~\ref{fig:fvszwaterfall} (a)), a component with a lower and decreasing frequency starts developing at $z \approx 5$\,m. %
Its amplitude dominates the one at the higher frequency at $z = 5.5$\,m, which causes the sudden jumps in frequency seen in Fig.~\ref{fig:fvsgz}. %
Then, the peak frequency follows $\fpe(z)$. 
Around this point the bunch train becomes significantly shorter (see loss of charge on Fig.~\ref{fig:charge_evolution}) and the driving of wakefields becomes dominated by microbunches early along the train. %
Because the relative dephasing between stretching wakefields and these early microbunches is smaller than with later ones that are totally defocused, their periodicity can better adjust to the local frequency at the expense of becoming shorter. %
Figure~\ref{fig:microbunch} (a) shows this effect and indicates that focusing wakefields can also bring charge back towards the axis, to regions that were previously depleted of charge, e.g., $\xi \approx 0.60, 0.85, 1.10$\,cm, with more charge later along the plasma than earlier. %

Figure~\ref{fig:fvszwaterfall} (b) for $g = +2$\,\%/m shows that in this case a second frequency appears at $z \approx 6$\,m with value $\fpe(z = 6\,\mr{m})$, which remains until the end of the plasma, but with a lower amplitude than that of the initial frequency value. %
In that case, after the modulation has saturated and the charge between microbunches has been completely depleted, the defocusing wakefields, which have a different frequency than the train, expel charge from the microbunches and modify the envelope of the train ($z = 10$\,m). %
Figure~\ref{fig:microbunch} (b) shows the resulting beating pattern. 
The experimental results also show two frequencies in the CTR spectrogram of some $g = +2$\,\%/m events~\cite{gradexp}. %
Streak camera images do not have sufficient signal to noise ratio to display the actual complexity of the frequency spectrum of numerical simulations. %
These figures show the intricacy of the frequency spectrum of the modulated bunch, resulting from the complex evolution of the bunch train and wakefields. %

The frequency analysis of the radial distribution after the plasma shows that detailed information can in general be retrieved from the experimental profiles and confirmed by simulation results, and even explained. %
This is possible because there is good agreement between experimental and simulation results, even though the SM process is an intricate one. %
The images of Fig.~\ref{fig:allprofiles} show the distribution of particles that drove wakefields towards the end of the plasma (with the added transverse evolution in vacuum between the plasma exit and the screen). %
They therefore also have some of the characteristics of the wakefields themselves, as seen in simulations. %
In that sense, they carry information similar to that obtained from various plasma density perturbation diagnostics such as interferometry, shadowgraphy~\cite{bib:shadowgraphy}, and Fourier domain holography \cite{bib:matlis}. %
This correspondence could be established with these diagnostics, as is the plan for later experiments. %

\section{Conclusion \label{sec:end}}

The detailed simulation and experimental results presented here regarding the charge in the microbunch train, its  distribution and  modulation frequency, as measured after the plasma as a function of linear plasma density gradients, are in excellent agreement with each other. %
They complement and explain our previously published results~\cite{gradexp} and show that many details of the self-modulation process are observed in the experiment. %
Density gradients change the phase velocity of the wakefields that we observe as a change in modulation frequency of the proton distributions measured along the plasma in simulations, and after the plasma in simulations and experiments. %
The effect of the density gradients is also to change the charge in the microbunch train and in the microbunches, as well as to change the amplitude of the wakefields. %
Time-resolved images of the proton bunch charge distribution contain information about the evolution of the self-modulation process along the plasma. %
For example, the modulation frequency of the defocused proton density distribution indicates the position along the plasma where protons left the wakefields. %
We will record similar data when optimizing a plasma density step~\cite{lotovdensitystep}, an extreme case of density gradient, for maximum plasma density modulation or maximum energy gain by externally injected electrons. %
This data will bring additional information about the effect of the density step on the proton bunch modulation. %

\FloatBarrier


\begin{acknowledgments}
This work was supported in parts by a Leverhulme Trust Research Project Grant RPG-2017-143 and by STFC (AWAKE-UK, Cockcroft Institute core, John Adams Institute core, and UCL consolidated grants), United Kingdom;
a Deutsche Forschungsgemeinschaft project grant PU 213-6/1 ``Three-dimensional quasi-static simulations of beam self-modulation for plasma wakefield acceleration''; 
the National Research Foundation of Korea (Nos.\ NRF-2016R1A5A1013277 and NRF-2020R1A2C1010835); 
the Portuguese FCT---Foundation for Science and Technology, through grants CERN/FIS-TEC/0032/2017, PTDC-FIS-PLA-2940-2014, UID/FIS/50010/2013 and SFRH/IF/01635/2015; 
the U.S.\ National Science Foundation under grant PHY-1903316; 
the Wolfgang Gentner Programme of the German Federal Ministry of Education and Research (grant no.\ 05E15CHA); 
M. Wing acknowledges the support of DESY, Hamburg. 
Support of the National Office for Research, Development and Innovation (NKFIH) under contract number 2019-2.1.6-NEMZ KI-2019-00004 and the support of the Wigner Datacenter Cloud facility through the Awakelaser project is acknowledged.
The work of V. Hafych has been supported by the European Union's Framework Programme for Research and Innovation Horizon 2020 (2014--2020) under the Marie Sklodowska-Curie Grant Agreement No.\ 765710.  
The AWAKE collaboration acknowledge the SPS team for their excellent proton delivery.
\end{acknowledgments}

\bibliography{gradsimbib.bib}

\end{document}